\documentstyle[12pt,amsmath,amsfonts]{article}
\def\note#1{}
\def\ij{\sum_{\stackrel{i,j=1}{\scriptscriptstyle i\neq j}}^N}

\begin{document}
\baselineskip 21pt
\vspace{1cm}
\begin{center}

{\Large\bf A note on the action-angle variables for
the rational  Calogero--Moser system }

\vskip 1cm

{\large {\bf Tomasz Brzezi\'nski \footnote{Lloyd's of London 
Tercentenary Fellow. On
leave from: Department of Theoretical Physics, University of 
\L\'od\'z, ul. Pomorska 149/153, 90-236 {\L}\'od\'z,
Poland. E-mail: tb10@york.ac.uk}}}

\vskip 0.1 cm
Department of Mathematics, University of York\\
Heslington, York YO10 5DD, England\\

\vskip 1cm

{\large {\bf Cezary Gonera}, {\bf Piotr Kosi\'nski} and {\bf Pawe{\l} Ma\'slanka}}

\vskip 0.1 cm

Department of Theoretical Physics II, University of {\L}\'od\'z,\\
ul. Pomorska 149/153, 90-236 {\L}\'od\'z, Poland\\

\end{center}
\vspace{1 cm}
\begin{abstract}
\begin{quote}
\noindent A relationship between the  action-angle variables and 
the canonical transformation  relating the 
rational Calogero--Moser system to the
free one is discussed.
\end{quote} 
\end{abstract}
\thispagestyle{empty}
\newpage 

\noindent The aim of this note is to answer the question of S.\ Ruijsenaars
\cite{Rui:pri} concerning the relationship between the action-angle
variables \cite{rui88} for the rational Calogero-Moser model
\cite{cal69} and the  equivalence of the latter to free particle systems described
explicitly with the help of $sl(2,\mathbb{R})$ dynamical symmetry in
\cite{BrzGon:equ}. 

We begin by recalling the construction of the canonical transformation
in \cite{BrzGon:equ}. This construction is based on the
observation that many features of the rational Calogero-Moser model 
with the Hamiltonian
$$
  H_{CM}= \sum_{i=1}^N \frac{p_{i}^{2}}{2}+\frac{g}{2} \sum_{i \neq j} 
  \frac{1}{(q_{i}-q_{j})^2},
 $$
where $p_i, q_i$ are the canonical variables, 
and $g$ is a coupling
constant, can be explained in terms of the dynamical 
$sl(2,\mathbb{R})$ symmetry. Consider the following four  
functions on the phase space:
$$
T_+=\frac{1}{\omega} H_{CM}, 
\quad T_-\equiv\omega\sum_{i=1}^N\frac{q_{i}^{2}}{2}, \quad 
T_0\equiv\frac{1}{2}\sum_{i=1}^Nq_i p_i, \quad 
\widetilde{T}_+ =\frac{1}{\omega}
    \sum_{i=1}^N\frac{p_{i}^{2}}{2},
$$
where $\omega\neq 0$ is a parameter. One easily checks that each of the sets 
$\{ T_+, T_-, T_0 \}$
and $\{ \widetilde{T}_+, T_-, T_0\}$ spans  the
$sl(2,\mathbb{R})$ Lie algebra with respect to the Poisson brackets,
i.e.,
$$
\{T_0, T_\pm\} = \pm T_{\pm}, \quad \{T_-,T_+\} =2T_0,
$$
and
$$
\{T_0,\widetilde{T}_+\} = \widetilde{T}_{+},\quad \{T_0,T_-\} = -T_{-}, \quad
\{T_-,\widetilde{T}_+\} =2T_0.
$$
These $sl(2,\mathbb{R})$ algebras act on the phase space in 
the standard way
by means of the Poisson brackets. The action can be integrated to the
symplectic action of the $SL(2,\mathbb{R})$ group. In the construction
of the transformation from the Calogero-Moser system to free particles
an important role is played by the following one-parameter family of
canonical transformations
\begin{eqnarray}
  && q_k\rightarrow \mbox{e}^{i\lambda T_1}\ast q_k\equiv 
     \sum_{n=0}^{\infty} \frac{(i\lambda)^n}{n!} 
     \left\{T_1,\dots\left\{T_1, q_k\right\}\dots\right\},\nonumber\\
  && p_k\rightarrow \mbox{e}^{i\lambda T_1}\ast p_k\equiv 
     \sum_{n=0}^{\infty} \frac{(i\lambda)^n}{n!} 
     \left\{T_1,\dots\left\{T_1, p_k\right\}\dots\right\}, \label{7}
\end{eqnarray}
where $T_1 = \frac{i}{2}(T_+ + T_-)$. Since $T_1=\frac{i}{2\omega}H_C$,
where
$$
H_C =\sum_{i=1}^N\left(\frac{p_{i}^{2}}{2}+
  \frac{\omega^2 q_{i}^{2}}{2}\right)+\frac{g}{2}\ij
  \frac{1}{(q_i-q_j)^2}
$$
 is the Hamiltonian of the Calogero model, the transformation (\ref{7})
can be viewed as the time evolution generated by the Calogero
Hamiltonian $H_C$, with the time $t=\lambda/2\omega$. On the other hand the
transformation (\ref{7}) is simply a rotation in the space spanned by
$T_0, T_\pm$ about the axis $T_1$ by an angle $\lambda$. Thus
for $\lambda=\pi$ (i.e., $t=\pi/2\omega$) we have
$
H_{CM} = \omega T_+\to \omega e^{i\pi T_1}*T_+ = \omega T_-$. Next we
can make a rotation in the space spanned by $T_0,T_-,\widetilde{T}_+$
about the axis $\widetilde{T}_1 = \frac{i}{2}(\widetilde{T}_++T_-)$
through the angle $-\pi$. In particular, this will rotate 
$T_-$ to $\widetilde{T}_+$. Since the latter is proportional to the
Hamiltonian of the free theory $H_0 = \sum_{i=1}^N \frac{p_i^2}{2}$, the
canonical transformation obtained by the combination of two rotations
transforms the rational Calogero-Moser model to the free particle
theory, i.e.,
$$
H_{CM}\to e^{-i\pi\widetilde{T}_1}*(e^{i\pi T_1}* H_{CM}) = H_0.
$$
Furthermore, this transformation sends the standard integrals of motion
of the Calogero-Moser model $\frac{1}{m}{\rm Tr}(L^m)$, $m=1,\ldots, N$,
 where 
$L$ is the Lax matrix, 
\begin{equation}
L_{jk} =
\delta_{jk} p_k +(1-\delta_{jk})\frac{ig}{q_j-q_k},
\label{lax}
\end{equation}
 to their free
counterparts (obtained by setting $g=0$). The same applies to the
functions ${\rm Tr}(QL^m)$, $m=1,\ldots, N$, with $Q={\rm diag}(q_1, q_2,\ldots,
q_N)$. 

The Ruijsenaars construction of the action-angle variables for the
Calogero-Moser model can be most clearly explained in terms of the
Hamiltonian reduction \cite{KazKos:...}. We now briefly recall how the
reduction procedure can be applied to the Calogero-Moser model
\cite{ols81}. One starts with the space of pairs $(A,B)$ of
$N\times N$ hermitian matrices. This space is equipped with the
symplectic form
\begin{equation}
\Omega = {\rm Tr}(dB\wedge dA).
\label{omega}
\end{equation}
The action of the unitary group $U(N)$,
\begin{equation}
U \in U(N) :\qquad (A,B)\to (UAU^\dagger, UBU^\dagger),
\label{unit}
\end{equation} 
preserves the form $\Omega$ in (\ref{omega}) and thus is a symplectic
action. The reduced phase space is obtained with the help of the
momentum map equation
$$
i[A,B] = g(I -{\bf v}^\dagger\otimes {\bf v}), \qquad {\bf v} =(1,1,\ldots, 1).
$$
Using the symplectic action of the group $U(N)$ in equation
(\ref{unit}),  one can fix a gauge in
which $A= {\rm diag}(q_1,q_2,\ldots, q_N)$. In this gauge $B$ is the Lax
matrix in equation (\ref{lax}), and $\Omega$ takes the standard form
$\Omega = \sum_{i=1}^N dp_i\wedge dq_i$. Thus we conclude that the
Calogero-Moser model can be obtained by the Hamiltonian reduction of a
simple dynamical system in $\Gamma$ defined by the Hamiltonian
$H=\frac{1}{2}{\rm Tr}B^2$.

On the other hand the symplectic transformation \cite{Mos:var} $A\to
\widetilde{A} = B$, $B\to \widetilde{B} = -A$ preserves the momentum
map. Following Ruijsenaars we can fix a gauge in which $\widetilde{A}
= B$ is diagonal, i.e., $\widetilde{A} = {\rm diag}(I_1,I_2,\ldots,
I_N)$. In this gauge the Hamiltonian $H = \frac{1}{2}{\rm Tr}B^2$ is $H
= \frac{1}{2}\sum_{i=1}^N I_i^2$. Clearly, the variables 
$I_1,\ldots, I_N$ are constants of motion
and together with the diagonal elements $-\phi_1,\ldots, -\phi_N$ of 
$\widetilde{B} = -A$ in this gauge, form the complete set of canonical
variables. Thus we conclude that $(\phi_i, I_i)$ are the action-angle
variables for the matrix model. 

It is now not difficult to relate this construction of action-angle
variables to that of the canonical map \cite{BrzGon:equ} recalled at the
beginning of this note. The action of the $sl(2,\mathbb{R})$ symmetry on
the reduced phase space can be lifted to $\Gamma$. Using the explicit form of the Poisson brackets induced by the
canonical form $\Omega$ (\ref{omega}), $\{ A_{ij},B_{kl}\} =
\delta_{il}\delta_{jk}$ one easily verifies that the
functions 
$$
t_+ = \frac{1}{2\omega}{\rm Tr}B^2, \quad t_-=\frac{\omega}{2} {\rm
Tr}A^2, \quad t_0 = \frac{1}{2} {\rm
Tr}AB ,
$$
generate the $sl(2,\mathbb{R})$ Lie algebra, i.e., 
$\{t_0, t_\pm\} = \pm t_{\pm}$, $\{t_-,t_+\} =2t_0$. 
The relationship between the 
actions of $sl(2,\mathbb{R})$ on the
unreduced and reduced phase spaces can be summarised in the following
commutative diagram:
$$
\begin{picture}(300,135)(10,10)
\put(10,10){\framebox(60,30){$(A,B)$}}
\put(40,95){\vector(0,-1){50}}
\put(45,65){$sl(2,\mathbb{R})$}
\put(10,100){\framebox(60,30){$(A,B)$}}
\put(75,115){\vector(1,0){100}}
\put(95,120){reduction}
\put(180,100){\framebox(60,30){$(q,p)$}}
\put(75,25){\vector(1,0){100}}
\put(95,30){reduction}
\put(180,10){\framebox(60,30){$(q,p)$}}
\put(210,95){\vector(0,-1){50}}
\put(215,65){$sl(2,\mathbb{R})$}
\end{picture} 
$$
Again using the explicit form of the
Poisson brackets one finds that $t_0, t_{\pm}$ act linearly on $A,B$.
This means that for any fixed $k,l$, $(A_{kl},B_{kl})$ is an
$sl(2,\mathbb{R})$ doublet. Therefore a general $sl(2,\mathbb{R})$
transformation of
$\Gamma$  can be represented as
\begin{equation}
\left(\begin{array}{c} A\\
\frac{1}{\omega} B\end{array}\right) \to 
\left(\begin{array}{cc} \alpha &\beta \\ \gamma &\delta
\end{array}\right)\left(\begin{array}{c} A\\
\frac{1}{\omega} B\end{array}\right)
, \qquad \alpha\delta - \beta\gamma =1.
\label{sl2}
\end{equation}
The transformation (\ref{sl2}) is a lift of the $sl(2,\mathbb{R})$
action on the reduced phase space. Thus, in particular, the lift of the
canonical transformation induced by $e^{i\pi T_1}$ (cf.\ equation
(\ref{7})), must be of the form (\ref{sl2}). We have
$$
\left(\begin{array}{c} A\\
\frac{1}{\omega} B\end{array}\right) \to e^{i\pi t_1}*\left(\begin{array}{c} A\\
\frac{1}{\omega} B\end{array}\right) = \left(\begin{array}{cc} 0 & 1
 \\ -1 & 0
\end{array}\right)\left(\begin{array}{c} A\\
\frac{1}{\omega} B\end{array}\right) = \left(\begin{array}{c} 
\frac{1}{\omega} B\\ -A\end{array}\right) .
$$
This shows that the Ruijsenaars procedure corresponds to the lifting of
the construction of the canonical mapping of the Calogero-Moser system
to free particles in \cite{BrzGon:equ}. One has to keep in mind,
however, that the diagonal elements of $B$ are viewed as momentum
variables in the Ruijsenaars approach while in \cite{BrzGon:equ} they
are proportional to the position variables. This explains the need for
the additional transformation $e^{-i\pi\widetilde{T}_1}$ which exchanges
the momentum and position variables (and kills the factor $\omega$). 

The reasoning presented above explains also in a straightforward way why the
functions ${\rm Tr}L^n = {\rm Tr}B^n$ and ${\rm Tr}(QL^n) = {\rm
Tr}(AB^n)$ are transformed to their free counterparts, while it is no
longer the case for  ${\rm Tr}(Q^mL^n)$, $m\geq 2$. The point is that
 the ${\rm Tr}B^n$
and ${\rm
Tr}(AB^n)$ depend only on the eigenvalues of $B$ and diagonal
elements of $A$ in the gauge in which $B$ is diagonal, while the 
${\rm Tr}(Q^mL^n)$, $m\geq
2$ depend on non-diagonal elements of $A$ too.

One can quantise the matrix theory on unreduced phase space $\Gamma$. Since the
action of $sl(2,\mathbb{R})$ is linear, it can easily be implemented on
the quantum level too. Then one can use the quantum Hamiltonian reduction
 \cite{poly91}  and carry the Ruijsenaars procedure over to the
quantum case (for a different approach see \cite{rui87}). At this point
the main advantage of the procedure producing  the symplectic map in 
\cite{BrzGon:equ} is that it can be immediately quantised.
\medskip

\noindent {\bf Acknowledgments}

\noindent   This research is supported by the
British Council grant WAR/992/147. The work of PK is supported by
the grant KBN 2 P03B 134 16.


\end{document}